\documentclass[journal]{IEEEtran}
\usepackage{cite}

\usepackage{graphicx}

\ifCLASSINFOpdf
\else
\fi
\begin{document}
%
\title{Variability Improvement by Interface Passivation and EOT Scaling of InGaAs Nanowire MOSFETs}
%
%
%

\author{Jiangjiang~J.~Gu,~\IEEEmembership{Student~Member,~IEEE,}
        Xinwei~Wang, Heng~Wu, \\Roy~G.~Gordon,
        and~Peide~D.~Ye,~\IEEEmembership{Fellow,~IEEE}
\thanks{This work was supported in part by Air Force Office of Scientific Research (AFOSR) monitored by Prof. James C. M. Hwang and in part by Semiconductor Research Corporation (SRC) Focus Center Research Program (FCRP) Materials, Structures, and Devices (MSD) Focus Center.}
\thanks{J. J. Gu, H. Wu, and P. D. Ye are with the Department
of Electrical and Computer Engineering, Purdue University, West Lafayette,
IN, 47907 USA e-mail: (yep@purdue.edu).}
\thanks{X. Wang, and R. G. Gordon are with the Department of Chemistry and Chemical Biology, Harvard University, Cambridge, MA, 02138 USA.}
}

\maketitle

\begin{abstract}
High performance InGaAs gate-all-around (GAA) nanowire MOSFETs with channel length ($L_{ch}$) down to 20nm have been fabricated by integrating a higher-\emph{k} LaAlO$_{3}$-based gate stack with an equivalent oxide thickness of 1.2nm. It is found that inserting an ultrathin (0.5nm) Al$_{2}$O$_{3}$ interfacial layer between higher-\emph{k} and InGaAs can significantly improve the interface quality and reduce device variation. As a result, a record low subthreshold swing of 63mV/dec has been demonstrated at sub-80nm $L_{ch}$ for the first time, making InGaAs GAA nanowire devices a strong candidate for future low-power transistors.
\end{abstract}

\begin{IEEEkeywords}
Variability, MOSFET, InGaAs, nanowire.
\end{IEEEkeywords}

%
\IEEEpeerreviewmaketitle

\section{Introduction}
%
%
%
%
\IEEEPARstart{I}{II-V} compound semiconductors have recently been explored as alternative channel materials for future CMOS technologies~\cite{AlamoNature2011}. In$_{x}$Ga$_{1-x}$As gate-all-around (GAA) nanowire MOSFETs fabricated using either bottom-up~\cite{ThelanderTED2008,TomiokaIEDM2011} or top-down technology~\cite{GuIEDM2011,GuIEDM2012,XueIEDM2012} are of particular interest due to their excellent electrostatic control. Although the improvement of on-state and off-state device metrics has been enabled by nanowire width ($W_{NW}$) scaling, the scalability of the devices in \cite{GuIEDM2011} is greatly limited by the large equivalent oxide thickness ($EOT$) of 4.5nm. Aggressive $EOT$ scaling is required to meet the stringent requirements on electrostatic control~\cite{RadosavljevicIEDM2011,EgardIEDM2011,GuIEDM2012}. It is shown recently that sub-1nm $EOT$ with good interface quality can be achieved by Al$_{2}$O$_{3}$ passivation on planar InGaAs devices~\cite{SuzukiAPL2012}. Considering the inherent 3D nature of the nanowire structure, whether such a gate stack technology can be successfully integrated in the InGaAs nanowire MOSFET fabrication process remains to be shown. In addition, the electron transport in the devices~\cite{GuIEDM2011} can be enhanced by increasing the Indium concentration in the InGaAs nanowire channel, which promises further on-state metrics improvements such as on-current ($I_{ON}$) and transconductance ($g_{m}$).

In this letter, we fabricated In$_{0.65}$Ga$_{0.35}$As GAA nanowire MOSFETs with atomic layer deposited (ALD) LaAlO$_{3}$-based gate stack ($EOT$=1.2nm). ALD LaAlO$_{3}$ is a promising gate dielectric for future 3D transistors because of its high dielectric constant (\emph{k}=16), precise thickness control, excellent uniformity and conformality~\cite{HuangIEDM2009}. The effect of ultra-thin Al$_{2}$O$_{3}$ insertion on the device on-state and off-state characteristics has been systematically studied. It is shown that Al$_{2}$O$_{3}$ insertion effectively passivates the LaAlO$_{3}$/InGaAs interface, leading to the improvement in both device scalability and variability. Record low subthreshold swing (\emph{SS}) of 63mV/dec has been achieved at sub-80nm $L_{ch}$, indicating excellent interface quality and gate electrostatic control. Detailed device variation analysis has been presented for the first time for InGaAs MOSFETs, which helps identify new manufacturing challenges for future logic devices with high mobility channels.

\section{Experiment}
\begin{figure}[h]
  \centering
  \includegraphics[width=0.4\textwidth]{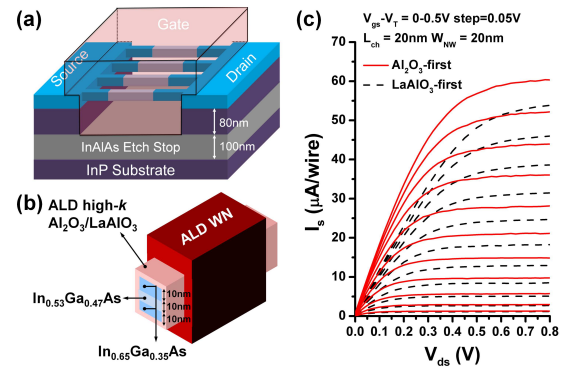}
  \caption{(a) Schematic diagram and (b) cross sectional view of InGaAs GAA nanowire MOSFETs with ALD Al$_{2}$O$_{3}$/LaAlO$_{3}$ gate stack. (c) Output characteristics (source current) of InGaAs GAA nanowire MOSFETs ($L_{ch}$=20nm) with Al$_{2}$O$_{3}$-first (solid line) and LaAlO$_{3}$-first (dashed line) gate stack.}
  \label{Figure1}
\end{figure}
Fig. \ref{Figure1}(a) and (b) show the schematic diagram and cross sectional view of InGaAs GAA nanowire MOSFETs fabricated in this work. The fabrication process is similar to that described in \cite{GuIEDM2011}. A HCl-based wet etch process was used to release the InGaAs nanowires with minimum $W_{NW}$ of 20nm. Each device had 4 nanowires in parallel as shown in Fig. \ref{Figure1}(a). Because of the relatively high etch selectivity between InAlAs and InP, an additional 100nm InAlAs etch stop layer was added under the 80nm InP sacrificial layer to improve the control of the nanowire release process. The InGaAs nanowire channel consists of one 10nm In$_{0.53}$Ga$_{0.47}$As layer sandwiched by two 10nm In$_{0.65}$Ga$_{0.35}$As layers shown in Fig. \ref{Figure1}(b), yielding a total nanowire height ($H_{NW}$) of 30nm. Here the heterostructure design ensures the high quality epitaxial layers grown by molecular beam epitaxy while maximizing the Indium concentration in the nanowire. A 0.5nm Al$_{2}$O$_{3}$, 4nm LaAlO$_{3}$, and 40nm WN high-k/metal gate stack were grown by ALD surrounding all facets of the nanowires. Two samples were fabricated in parallel with only the sequence of the Al$_{2}$O$_{3}$ and LaAlO$_{3}$ growth deliberately switched. Both samples were treated with 10\% (NH$_{4}$)$_{2}$S, and then transferred into the ALD chamber within 1 minute of air break. Since the Al$_{2}$O$_{3}$-first and LaAlO$_{3}$-first sample had the same $EOT$ of 1.2nm and underwent the same process flow, the difference of device performance can be ascribed to the effect of the Al$_{2}$O$_{3}$ passivation. All other fabrication details can be found in \cite{GuIEDM2011}. In this letter, the channel length ($L_{ch}$) is defined as the width of the electron beam resist in the source/drain implantation process and has been verified by scanning electron microscopy.

\section{Results and discussion}
\begin{figure}[h]
  \centering
  \includegraphics[width=0.4\textwidth]{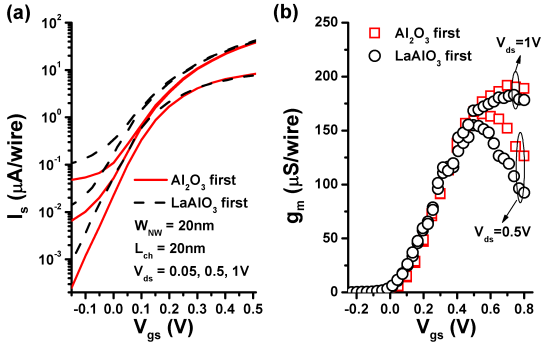}
  \caption{(a) Transfer characteristics (source current) at $V_{ds}$=0.05, 0.5, and 1V (b) $g_{m}$-$V_{gs}$ of Al$_{2}$O$_{3}$-first and LaAlO$_{3}$-first InGaAs GAA nanowire MOSFETs with $L_{ch}=$20nm.}
  \label{Figure2}
\end{figure}
Fig. \ref{Figure1}(c) shows the output characteristics of two representative Al$_{2}$O$_{3}$-first and LaAlO$_{3}$-first InGaAs GAA nanowire MOSFETs with $L_{ch}=20nm$. Fig. \ref{Figure2}(a) and (b) show the transfer characteristics and transconductance of the same devices. Due to the large junction leakage current in the drain, the source current $I_{s}$ is shown in the current-voltage characteristics and used to calculate $I_{ON}$ and $g_{m}$. The Al$_{2}$O$_{3}$-first device shows higher $I_{ON}=57\mu$A/wire at $V_{DD}=V_{ds}=V_{gs}-V_{T}=0.5V$ and peak transconductance $g_{m,max}=165\mu$S/wire at $V_{ds}=0.5V$, compared to 48$\mu$A/wire and 155$\mu$S/wire for the LaAlO$_{3}$-first device. Both devices operate in enhancement-mode, with a linearly extrapolated $V_{T}$ of 0.14V and 0.11V, respectively. For the off-state performance, the Al$_{2}$O$_{3}$-first device shows a $SS$ of 75mV/dec and $DIBL$ of 40mV/V, while the LaAlO$_{3}$-first device shows higher $SS$ of 80mV/V and higher $DIBL$ of 73mV/V.
\begin{figure}[h]
  \centering
  \includegraphics[width=0.4\textwidth]{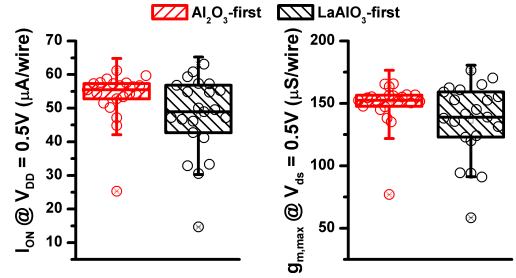}
  \caption{$I_{ON}$ and peak $g_{m}$ box plots of Al$_{2}$O$_{3}$-first and LaAlO$_{3}$-first devices with $L_{ch}=20nm$ and $W_{NW}=20nm$ at $V_{DD}=0.5V$.}
  \label{Figure3}
\end{figure}
To study the statistical distribution of the on-state metrics, the box plots for $I_{ON}$ and $g_{m,max}$ at $V_{DD}=0.5V$ are shown in Fig. \ref{Figure3}. The box plots include measurements from all 50 devices with $L_{ch}$ of 20nm and $W_{NW}$ of 20nm. Although only a 12\% (10\%) increase in mean $I_{ON}$ ($g_{m,max}$) is observed for the devices with Al$_{2}$O$_{3}$ insertion, a 54\% (64\%) reduction in standard deviation of $I_{ON}$ ($g_{m,max}$) is obtained on the Al$_{2}$O$_{3}$-first devices, indicating a significant improvement in device variation by effective passivation of interface traps. The $I_{ON}$ variation is impacted by several variation sources including parasitic resistance, effective mobility and $V_{T}$ variation~\cite{MatsukawaTED2012}, all of which are sensitive to the interface quality of the high-\emph{k}/InGaAs nanowire surface.

\begin{figure}[h]
  \centering
  \includegraphics[width=0.4\textwidth]{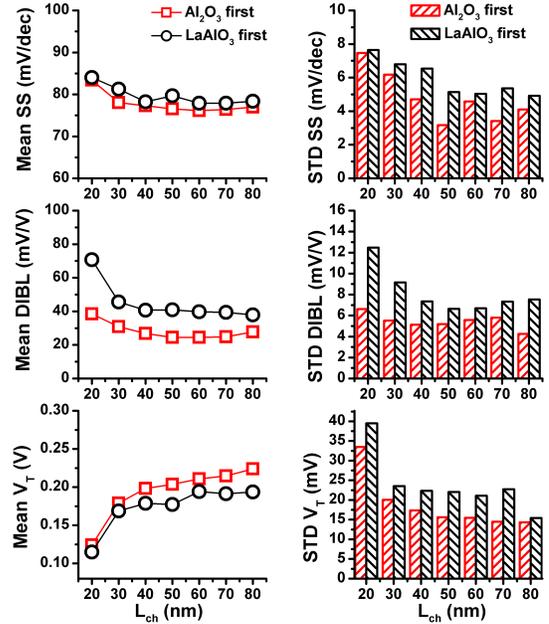}
  \caption{Scaling metrics of $SS$, $DIBL$ and $V_{T}$ and their standard deviations (STDs) for Al$_{2}$O$_{3}$-first and LaAlO$_{3}$-first InGaAs GAA nanowire MOSFETs with $W_{NW}=20nm$}
  \label{Figure4}
\end{figure}

To further investigate the scalability and off-state performance variability, the averages and standard deviations of $SS$, $DIBL$ and $V_{T}$ as a function of $L_{ch}$ are shown in Fig. \ref{Figure4} for Al$_{2}$O$_{3}$-first and LaAlO$_{3}$-first devices with $W_{NW}=20nm$. The $SS$ and $DIBL$ remain almost constant with $L_{ch}$ scaling down to 50nm for both samples. This indicates that the current GAA structure with 1.2nm $EOT$ has yielded a very small geometric screening length and the devices show excellent resistance to short channel effects. Average $SS=76$mV/dec and $DIBL=25$mV/V are obtained for Al$_{2}$O$_{3}$-first devices with $L_{ch}$ between 50 and 80nm, compared to 79mV/dec and 39mV/V for the LaAlO$_{3}$-first devices, indicating a reduction of interface trap density ($D_{it}$) with Al$_{2}$O$_{3}$ passivation. A small increase in $V_{T}$ is also observed for the Al$_{2}$O$_{3}$-first sample, which is ascribed to the reduction in negative donor-type charges at the interface.
Furthermore, larger standard deviations of $SS$, $DIBL$ and $V_{T}$ are observed for devices without Al$_{2}$O$_{3}$ insertion at all $L_{ch}$, indicating that the relatively low interface quality of the LaAlO$_{3}$-first devices introduced additional device variation. It is also shown that the off-state performance variation increases as $L_{ch}$ scales below 50nm, which is ascribed to the reduction in electrostatic control.

\begin{figure}[h]
  \centering
  \includegraphics[width=0.4\textwidth]{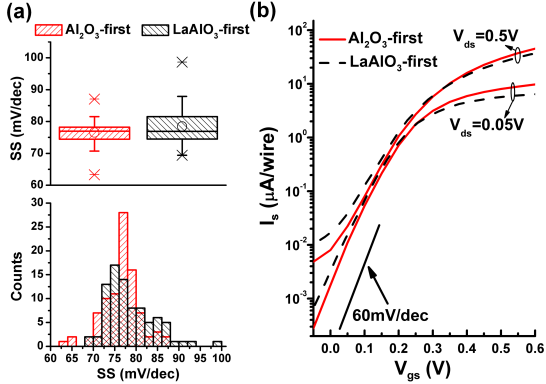}
  \caption{(a) $SS$ box plot and histogram for all Al$_{2}$O$_{3}$-first and LaAlO$_{3}$-first devices with $L_{ch}$ between 50$-$80nm and $W_{NW}$ of 20nm. (b) Transfer characteristic (source current) of a Al$_{2}$O$_{3}$-first and a LaAlO$_{3}$-first InGaAs GAA nanowire MOSFET with lowest SS of 63mV/dec and 69mV/dec, respectively.}
  \label{Figure5}
\end{figure}
Fig. \ref{Figure5}(a) show the box plot and histogram of $SS$ measured from all the Al$_{2}$O$_{3}$-first and LaAlO$_{3}$-first devices with $L_{ch}$ between 50$-$80nm and $W_{NW}$ of 20nm. Although the average $SS$ for Al$_{2}$O$_{3}$-first devices is only 1.9mV/dec lower than LaAlO$_{3}$-first devices, 25\% and 46\% reduction in standard deviation and interquartile range has been obtained on Al$_{2}$O$_{3}$-first devices, indicating the effectiveness of Al$_{2}$O$_{3}$ passivation. Since these devices are immune to short channel effects, the $SS$ is dominated by $D_{it}$. Therefore, $D_{it}$ can be estimated from $SS$ using the following equation,
\begin{equation}\label{SSequation}
SS=\frac{60}{300}T(1+(\frac{qD_{it}}{C_{ox}}))mV/dec
\end{equation}
where $T$ is the temperature in Kelvin, $q$ is the electronic charge, and $C_{ox}$ is the oxide capacitance. 90\% of the devices with Al$_{2}$O$_{3}$ insertion show $SS$ between $66.0-83.3$mV/dec, corresponding to a $D_{it}$ between $1.80\times10^{12}-6.98\times10^{12}$cm$^{-2}$eV$^{-1}$. Fig. \ref{Figure5}(b) shows the transfer characteristics of an 80nm $L_{ch}$ hero Al$_{2}$O$_{3}$-first device and a 60nm $L_{ch}$ hero LaAlO$_{3}$-first device with the lowest $SS=63$mV/dec and 69mV/dec, respectively. The estimated $D_{it}$ for these two devices are $8.98\times10^{11}$ and $2.69\times10^{12}$cm$^{-2}$eV$^{-1}$. The near-ideal $SS$ is achieved because of the surface area of the nanowires, aggressive \emph{EOT} scaling, and effective interface passivation.

\section{Conclusion}
 InGaAs GAA nanowire MOSFETs with $L_{ch}$ down to 20nm and $EOT$ down to 1.2nm have been demonstrated, showing excellent gate electrostatic control. The insertion of an ultra-thin 0.5nm Al$_{2}$O$_{3}$ between LaAlO$_{3}$/InGaAs interface has shown to effectively improve the scalability and variability of the devices. Near-60mV/dec $SS$ is achieved on InGaAs nanowires with scaled $EOT$ and effective interface passivation. The InGaAs GAA nanowire MOSFET is a promising candidate for low-power logic applications beyond 10nm.


%



\section*{Acknowledgment}
The authors would like to thank A. T. Neal, M. S. Lundstrom, D. A. Antoniadis, and J. A. del alamo for the valuable discussions.

\ifCLASSOPTIONcaptionsoff
  \newpage
\fi




\bibliographystyle{IEEEtran}
\end{document}